\documentclass[conference]{IEEEtran}
\IEEEoverridecommandlockouts
\usepackage{cite}
\usepackage{booktabs}
\usepackage{upgreek}
\usepackage{amsmath,amssymb,amsfonts}
\usepackage{algorithmic}
\usepackage{graphicx}
\usepackage{textcomp}
\usepackage{xcolor}
\usepackage[table]{xcolor}
\usepackage{hyperref}
\usepackage{makecell}
\usepackage{multirow}
\allowdisplaybreaks
\def\BibTeX{{\rm B\kern-.05em{\sc i\kern-.025em b}\kern-.08em
    T\kern-.1667em\lower.7ex\hbox{E}\kern-.125emX}}
\begin{document}

\title{Analyzing the Impact of Demand Response on Short-Circuit Current via a Unit Commitment Model
\thanks{This work was supported by MICIU/AEI/10.13039/501100011033 and ERDF/EU under grant PID2023-150401OA-C22, as well as by the Madrid Government (Comunidad de Madrid-Spain) under the Multiannual Agreement 2023-2026 with Universidad Politécnica de Madrid, ``Line A - Emerging PIs''. The work of Peng Wang was also supported by China Scholarship Council under grant 202408500065.}
}

\author{
\IEEEauthorblockN{Peng Wang\IEEEauthorrefmark{1}, Zhengmao Li\IEEEauthorrefmark{2} and Luis Badesa\IEEEauthorrefmark{1}}
\IEEEauthorblockA{\IEEEauthorrefmark{1}\textit{School of Industrial Engineering and Design (ETSIDI), Technical University of Madrid (UPM), Spain}
\\peng.wang@alumnos.upm.es, luis.badesa@upm.es}
\IEEEauthorblockA{\IEEEauthorrefmark{2}\textit{School of Electrical and Electronic Engineering, Aalto University, Finland}
\\{zhengmao.li@aalto.fi}}
}

\maketitle

\begin{abstract}
In low-carbon grids, system flexibility can be enhanced through mechanisms such as Demand Response (DR), enabling the efficient utilization of renewable energy. However, as Synchronous Generators (SGs) are being replaced by renewable energy sources characterized by Inverter-Based Resources (IBR), system stability is severely affected. Due to the limited overload capability of IBRs, their Short-Circuit Current (SCC) contribution is much smaller than that of SGs. As a result, protection devices may fail to trip during faults. Consequently, the remaining SGs play a key role in providing sufficient SCC. Since the commitment of SGs is closely related to system loading conditions, DR can indirectly affect their SCC provision, a relationship that has not yet been investigated in the literature. Therefore, this paper incorporates both DR and SCC constraints into a unit commitment problem and conducts case studies on an IEEE 30-bus system. The results show that although DR can reduce total costs by adjusting power demand, it may also lead to inadequate SCC levels. Nevertheless, when flexible loads are properly coordinated with SCC requirements, the total cost increases by only 0.3\%, which is significantly lower than the cost of system dispatch without DR. This demonstrates that DR can facilitate stable system operation in a cost-effective manner.
\end{abstract}


\section*{Nomenclature}
\addcontentsline{toc}{section}{Nomenclature}


\vspace{-0.2cm}
\subsection*{Indices and Sets}
\begin{IEEEdescription}
    \item[$c,\mathcal{C}$] \qquad Index, Set of IBR
    \item[$g,\mathcal{G}$] \qquad Index, Set of SGs
    \item[$t, T$]  \qquad Index, Set of time periods
\end{IEEEdescription}

\vspace{-0.4cm}
\subsection*{Constants and Parameters}
\begin{IEEEdescription}
    \item[$\textrm{c}^{\textrm{SL,in}},\textrm{c}^{\textrm{SL,out}}$] \qquad ~Compensation for adjustment of SL (\texteuro/MWh)
    \item[$\textrm{c}^\textrm{st}_g,\textrm{c}^\textrm{sh}_g$] Startup/shutdown costs of SGs (\texteuro/h) 
    \item[$\textrm{c}_n^\textrm{IL}$] Compensation for adjustment of type-$n$ IL (\texteuro/MWh)
    \item[$\textrm{c}_g^\textrm{nl}$]  No-load costs of SGs (\texteuro/h)
    \item[$\textrm{c}_g^\textrm{m}$]  Marginal power generation costs of SGs (\texteuro/MWh)  
    \item[$\textrm{I}_{b_{\textrm{lim}}}$] Safe threshold for SCC level (p.u.)
    \item[$\textrm{P}^\textrm{base}_t$] System demand before the DR (MWh) 
    \item[$\textrm{P}_c^\textrm{max}$] Available power of IBR (MW)
    \item[$\textrm{P}_g^\textrm{min},\textrm{P}_g^\textrm{max}$] \qquad  \parbox[t]{.9\linewidth} {Bounds for output of SGs (MW)} 
    \item[$\upbeta_{\{ \textrm{1,2,3,4} \}}$]  \qquad  \parbox[t]{.9\linewidth} {Ratio of flexible load}
    \item[$\uplambda_t$] Energy price (\texteuro/MWh)   
    \item[$\upalpha_{c,t}$] Capacity percentage of wind turbines 
\end{IEEEdescription}

\vspace{-0.4cm}
\subsection*{Variables}
\begin{IEEEdescription}
    \item[$C_t^\textrm{E}$] Final payment for consumers (\texteuro)
    \item[$C_{g,t}^\textrm{st}, C_{g,t}^\textrm{sh}$] \quad~ \parbox[t]{.9\linewidth} {Startup/shutdown costs incurred by SGs at time period $t$ (\texteuro/h)}
    \item[$P^\textrm{DR}_t$] Power demand after the DR (MWh)
    \item[$P_{g,t}$]  Power output of SGs (MW)
    \item[$P_{c,t}$]  Power output of wind turbines (MW)
    \item[$P_{n,t}^{\textrm{curt}}$] Curtailment of type-$n$ IL (MWh)
    \item[$P_t^{\textrm{in}},P_t^{\textrm{out}}$] \quad SL that shift in/out at period $t$ (MWh)
    \item[$z^{\textrm{in}}_t,z^{\textrm{out}}_t$] \quad Binary variable, response decision of SL
    \item[$u_{g,t}$] Binary variable, commitment of SG 
    \item[$\eta_{m,t}$] Binary variable, product of two SGs' commitments
\end{IEEEdescription}

\section{Introduction}
With the continuous increase in the penetration of renewable energy such as wind and solar power, the inherent variability of their output has posed challenges to the stability, flexibility of low-carbon power systems \cite{denholm2011grid}. Traditional generation-side regulation methods, such as ramping or startup/shutdown scheduling of thermal units, are often costly and less fit for high-renewable systems \cite{kubik2015increasing}. Consequently, enhancing system flexibility on the demand side has become essential for achieving reliable and economical grid operation \cite{renewables2021analysis,palensky2011demand}.

In this context, Demand Response (DR), which serves as a key mechanism for mobilizing end-users to balance supply and demand, can guide consumers to proactively adjust their electricity consumption behaviors in response to price signals or incentive mechanisms \cite{albadi2008summary}, thereby enabling an interactive relationship between users and the grid and improving system flexibility. Generally, DR programs are categorized into two types: price-based DR and incentive-based DR. The former motivates users to modify their electricity consumptions over different time periods through energy price signals, such as time-of-use pricing \cite{wang2023peak}, whereas the latter encourages users to adjust their power usage when needed through contractual obligations or compensation mechanisms, such as peak clipping or interruptible loads \cite{cappers2010demand}. Existing studies have demonstrated that implementing DR can yield benefits, such as reducing system costs, achieving peak shaving, and facilitating the efficient utilization of renewable energy \cite{dong2023values,pakbin2025optimized,mcpherson2020demand}.

Meanwhile, the large-scale integration of renewable energy into the grid has inevitably led to the gradual replacement of conventional Synchronous Generators (SGs) with Inverter-Based Resources (IBR) such as wind turbines. Unlike SGs, which are electro-mechanically coupled to the grid and inherently provide stability characteristics, IBR interface through power electronic converters that operate under control algorithms. As a result, they exhibit fundamentally different dynamic behaviors and grid-supporting capabilities \cite{chaudhuri2024rebalancing}.

From a protection standpoint, one of the major challenges associated with the increasing share of IBR is the decline in Short-Circuit Current (SCC) at system buses. Conventional protective relays are designed to detect and isolate faults based on sufficiently high fault currents; however, IBR inherently have limited overcurrent capability. This results in fault currents that are typically only 1.1–1.2 times their rated current, much lower than the 5–8 times seen in SGs \cite{jia2017review,chu2021short}. As a result, those devices may fail to timely operate, posing risks to fault isolation and system stability. The commitment of SGs that are still in service therefore plays a vital role in this issue.

It is evident that, by encouraging consumers to shift or curtail their energy usage, DR can effectively reshape the system load curve, thereby influencing the commitment of SGs. Since SGs are the primary contributors to the SCC, the implementation of DR may reduce the overall fault current capability of the grid when demand reduction occurs. To investigate this potential impact, a Unit Commitment (UC) model is developed to evaluate the influence of an incentive-based DR mechanism on system-wide SCC levels.

The remainder of this paper is organized as follows: Section \ref{Methodologies and Models} introduces the UC model including constraints of SCC and DR. Section \ref{Case Studies} includes case studies that analyze the impact of DR on the SCC level. Finally, Section \ref{Conclusion} concludes the paper and outlines future research.

\vspace{-0.1cm}
\section{Methodologies and Models}\label{Methodologies and Models}
This section introduces the UC model with constraints of SCC and DR. It should be noted that only three-phase nodal short-circuit faults are considered in this work.

\vspace{-0.1cm}
\subsection{Models of Short-Circuit Current Constraints}\label{Short Circuit Current Constraints}
For a power system incorporating SGs and IBR, the SCC at bus $b$ due to their current injections can be expressed as \cite{chu2021short}:
\begin{equation}
     I_{b_\textrm{SC}}= \frac{\sum_{g\in\mathcal{G}}Z_{b\Psi(g)}\textrm{I}_gu_g+\sum_{c\in\mathcal{C}}Z_{b\Phi(c)}\textrm{I}_c\upalpha_c}{Z_{bb}}  \label{eq:define_of_SCC_constraints}  
\end{equation}
where $\textrm{I}_g$ and $\textrm{I}_c$ are the short-circuit injections from SGs and IBR, respectively. $Z_{b\Psi(g)}$ indicates the impedance between bus $b$ and bus of SG $g$. This constraint involves the inverse operation between impedance and admittance matrices of the system, which is associated with the SGs' commitments.

To eliminate the need for matrix inversion during system dispatch, an approximation method can be used to obtain the linear SCC expression \cite{chu2021short}. It is an optimization-based classification procedure, comprising of an enumeration of system operating points and a minimization problem that leads approximated SCC values ($I_{b_\textrm{L}}$) to fit real ones ($I_{b_\textrm{SC}}$). For brevity, the detailed approximation process is not elaborated.

Subsequently, the approximated SCC at bus $b$ is given by:
\begin{subequations} \label{eq:SCC_constraints}   
\begin{align}
& I_{b_\textrm{L}} = \sum_{g}\textrm{k}_{bg}u_{g} +\sum_{c}\textrm{k}_{bc}\upalpha_{c} +\sum_{m}\textrm{k}_{bm}\eta_{m} \ge \textrm{I}_{b_{\textrm{lim}}} \label{eq:define_SCC} \\
& \eta_m=u_{g_\textrm{1}}u_{g_\textrm{2}} \in \{0, 1\}, \quad \textrm{s.t.}~\{g_\textrm{1},g_\textrm{2}\}=m    \label{eq:interactions_pair_SGs_1}         \\ 
& m\in\mathcal{M}=\{g_{\textrm{1}},g_{\textrm{2}}\mid \forall g_{\textrm{1}}, g_{\textrm{2}}\in\mathcal{G}\}    \label{eq:interactions_pair_SGs_2}   \\
& \eta_{m}\leq u_{g_\textrm{1}},~ \eta_{m}\leq u_{g_\textrm{2}},~ \eta_{m} \geq u_{g_\textrm{1}}+u_{g_\textrm{2}}-1 \label{eq:MC_linear}
    \end{align}
\end{subequations}
where \eqref{eq:define_SCC} confines the SCC level to be higher than $\textrm{I}_{b_\textrm{lim}}$ for a secure system operation. Eqs.~\eqref{eq:interactions_pair_SGs_1}–\eqref{eq:interactions_pair_SGs_2} capture the nonlinear terms standing for the simultaneous current injections of any pair of SGs. The coefficients $\{\textrm{k}_{bg},\textrm{k}_{bc},\textrm{k}_{bm}\}$ are optimized by the approximation process. The term `$\eta_m$' that represents the product of two binary variables is reformulated by McCormick envelopes, which are expressed via auxiliary equations~\eqref{eq:MC_linear}.

\subsection{Models of Demand Response}
In this work, DR is modeled through an incentive-based mechanism, which considers both Interruptible Load (IL) and Shiftable Load (SL) \cite{zhou2017strategic}. It is assumed that users adjust their power consumption (with the adjusted range decided by the original demand and corresponding flexible loads' ratio `$\upbeta_{\{\cdot\}}$') effectively upon receiving defined compensations.

\subsubsection{Interruptible Load}
The IL typically involves signing an agreement with power companies to reduce electricity consumption in exchange for financial compensation. The response of IL is modeled as:
\begin{equation} \label{eq:IL_model}
 0 \leq P_{n,t}^{\textrm{curt}} \leq \upbeta_{\textrm{1},n} \textrm{P}^{\textrm{base}}_t,~ P_{n,t-1}^{\textrm{curt}} + P_{n,t}^{\textrm{curt}} \leq \upbeta_{\textrm{2}} \textrm{P}^{\textrm{base}}_t
\end{equation}
where \eqref{eq:IL_model} limits IL curtailment to an allowable range over both a single period and a continuous time interval, respectively.

\subsubsection{Shiftable Load}
The SL is characterized by flexible consumption scheduling, i.e., the load can be shifted from peak to off-peak periods. The corresponding models are given as:
\begin{subequations} \label{eq:SL_model}
\begin{align}
& \qquad \qquad \qquad z_t^{\textrm{in}}  +  z_t^{\textrm{out}} \leq 1      \label{eq:SL_model_1} \\
& 0 \leq P_t^{\textrm{in}} \leq \upbeta_{\textrm{3}} \textrm{P}^{\textrm{base}}_t z_t^{\textrm{in}}, \quad 0 \leq P_t^{\textrm{out}} \leq \upbeta_{\textrm{4}} \textrm{P}^{\textrm{base}}_t z_t^{\textrm{out}} \label{eq:SL_model_3} \\
& \qquad \qquad \quad \sum_t P_t^{\textrm{in}} - \sum_t P_t^{\textrm{out}} =0 \label{eq:SL_model_4} 
\end{align}
\end{subequations}
where \eqref{eq:SL_model_1} specifies that only load shifting-out or shifting-in is allowed within a single time period. Eq.~\eqref{eq:SL_model_3} constrains the adjustment volume of SL. Eq.~\eqref{eq:SL_model_4} ensures that the total shifted amount is conserved.

Based on the above DR models, the energy demand that the generating units eventually need to meet, and the fee paid by the users can be respectively expressed by:
\begin{subequations}\label{eq:DR_quantity_fee}
\begin{align}
   & P^\textrm{DR}_t = \textrm{P}^\textrm{base}_t - P^\textrm{out}_t + P^\textrm{in}_t - \sum_n P_{n,t}^\textrm{curt} \label{eq:Demand_after_DR} \\ 
   & C_t^\textrm{E}  =  \biggl(  \uplambda_t P^\textrm{DR}_t \hspace{-0.1cm} - \sum_n \textrm{c}_n^\textrm{IL} P_{n,t}^\textrm{curt} - \textrm{c}^\textrm{SL,in} P_{t}^\textrm{in}  - \textrm{c}^\textrm{SL,out} P_{t}^\textrm{out}  \biggl) \label{eq:DR_compensation}
    \end{align}
\end{subequations}
where \eqref{eq:Demand_after_DR} represents the system demand after DR. Eq.~\eqref{eq:DR_compensation} is the total payment for consumers, comprising the energy cost and the compensation for adjustments of power consumption.

\vspace{-0.1cm}
\subsection{SCC Constrained UC Model with DR}
The objective of this UC is to minimize the total cost, that is, the sum of the system operating cost and the energy payments borne by users. With the inclusion of SCC constraints and the flexible load through DR, the UC is established as follows:
\begin{subequations}\label{eq:primal_model}
\vspace{-0.2cm}
\begin{align}
    \min \ & \hspace{-0.1cm}\sum_{t} \Biggl[
     \sum_g \biggl(\textrm{c}_{g}^\textrm{nl} u_{g,t} +\textrm{c}_{g}^\textrm{m} P_{g,t}+ C_{g,t}^\textrm{st} +  C_{g,t}^\textrm{sh} \biggr) +C_t^\textrm{E} \Biggr] \label{eq:primal_obj}\\
    \text{s.t.} \quad
    & \text{SCC constraints}, \quad \forall b,m,t~\eqref{eq:SCC_constraints} \nonumber\\
    & \text{DR constraints}, \quad \forall n,t~\eqref{eq:IL_model}, \eqref{eq:SL_model}, \eqref{eq:DR_quantity_fee} \nonumber\\
    & \sum_{g}P_{g,t}+ \sum_{c}P_{c,t} = P^\textrm{DR}_t, \quad \forall g,c,t \label{eq:primal_cons_power_balance}\\
    & u_{g,t} \textrm{P}_{g}^\textrm{min} \leq P_{g,t} \leq u_{g,t} \textrm{P}_{g}^\textrm{max}, \quad \forall g,t \label{eq:primal_cons_SG_output}\\
    & C_{g,t}^\textrm{st} \ge 0, \quad C_{g,t}^\textrm{sh} \ge 0, \quad \forall g,t \label{eq:primal_cons_positive}\\
    & C_{g,t}^\textrm{st} \ge (u_{g,t}-u_{g,(t-1)})\textrm{c}^\textrm{st}_{g}, \quad \forall g,t \label{eq:primal_cons_st_cost_lb}\\
    & C_{g,t}^\textrm{sh} \ge (u_{g,(t-1)}-u_{g,t})\textrm{c}^\textrm{sh}_{g}, \quad \forall g,t \label{eq:primal_cons_sh_cost_lb}\\
    & 0 \leq P_{c,t} \leq \upalpha_{c,t}\textrm{P}^{\textrm{max}}_{c}, \quad \forall c,t  \label{eq:primal_cons_wt}\\
    & u_{g,t} \in \{0,1\}, \quad \forall g,t \label{eq:binary_SGs}
\end{align}
\end{subequations}
where the system operating costs are represented in \eqref{eq:primal_obj}, including the no-load, marginal generation and startup/shutdown costs of SGs. Eq.~\eqref{eq:primal_cons_power_balance} is the supply-demand power balance. Eq.~\eqref{eq:primal_cons_SG_output} bounds output of SGs. Eqs.~\eqref{eq:primal_cons_positive}-\eqref{eq:primal_cons_sh_cost_lb} define startup/shutdown costs of SGs. Eq.~\eqref{eq:primal_cons_wt} gives generation limits for IBR. Eq.~\eqref{eq:binary_SGs} specifies the binary nature of UC decisions.

\vspace{-0.1cm}
\section{Case Studies}\label{Case Studies}

\begin{figure}[t]
\centering
\includegraphics[width=0.6\columnwidth]{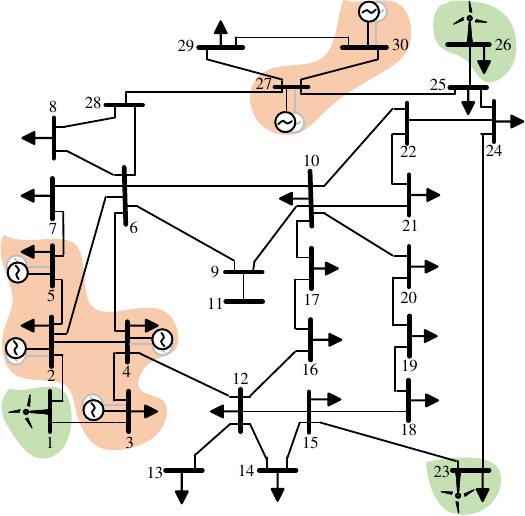}
\vspace{-0.2cm}
\caption{Modified IEEE 30-bus system.}
\label{ieee_system}
\vspace{-0.2cm}
\end{figure}   

\begin{table}[t]
\centering
\caption{Operating Parameters of Synchronous Generators}
\vspace{-0.2cm}
\setlength{\tabcolsep}{6.3pt}  
{\fontsize{8pt}{10pt}\selectfont
\begin{tabular}{lcccccc}
\toprule
Bus & 2 & 3 & 4 & 5 & 27 & 30 \\
\midrule
\(\textrm{c}_{g}^\textrm{nl}\) (\texteuro/h) & 1,743 & 1,501 & 1,376 & 1,093 & 990 & 857 \\
\(\textrm{c}_{g_1}^\textrm{m}\) (\texteuro/MWh) & 6.20 & 7.10 & 10.47 & 12.28 & 13.53 & 15.36 \\
\(\textrm{c}_{g_2}^\textrm{m}\) (\texteuro/MWh) & 7.07 & 8.72 & 11.49 & 12.84 & 14.60 & 15.02 \\
\(\textrm{c}_g^\textrm{st}\) (\texteuro/h) & 20,000 & 12,500 & 9,250 & 7,200 & 5,500 & 3,100 \\
\(\textrm{c}_g^\textrm{sh}\) (\texteuro/h) & 5,000 & 2,850 & 1,850 & 1,440 & 1,200 & 1,000 \\
\(\textrm{P}_g^\textrm{min}\) (MW) & 658 & 576 & 302 & 133 & 130 & 58 \\
\(\textrm{P}_g^\textrm{max}\) (MW) & 1,317 & 1,152 & 756 & 667 & 650 & 576 \\
\(u_{g,0}\) & 1 & 1 & 1 & 1 & 1 & 0 \\
\bottomrule
\end{tabular}
}
\label{table:SGs_para}
\vspace{-0.4cm}
\end{table}

\begin{table}[t!]
\centering
\caption{Parameters Associated with Demand Response}
\vspace{-0.2cm}
\setlength{\tabcolsep}{11pt}  
{\fontsize{8pt}{10pt}\selectfont
\begin{tabular}{lc}
\toprule
Parameters & Values \\
\midrule
\(\textrm{c}_{\textrm{I}}^{\textrm{IL}},\textrm{c}_{\textrm{II}}^{\textrm{IL}},\textrm{c}_{\textrm{III}}^{\textrm{IL}}\) (\texteuro/MWh) & 50, 70, 100   \\
\(\textrm{c}^{\textrm{SL,in}},\textrm{c}^{\textrm{SL,out}}\) (\texteuro/MWh) & 20, 30 \\
\(\upbeta_{\textrm{1,I}},\upbeta_{\textrm{1,II}},\upbeta_{\textrm{1,III}}\)  & 0.1, 0.08, 0.05  \\
\(\upbeta_{\textrm{2}},\upbeta_{\textrm{3}},\upbeta_{\textrm{4}}\)  & 0.2, 0.12, 0.12  \\
\bottomrule
\end{tabular}
}
\label{table:Parameters Associated with Demand Response}
\vspace{-0.4cm}
\end{table}

\begin{table}[!t]
\centering
\caption{ Errors in Approximation for SCC Constraints }
\vspace{-0.2cm}
{\fontsize{8pt}{10pt}\selectfont
\setlength{\tabcolsep}{6pt}
\begin{tabular}{c cc cc}
\toprule
\multirow{2}{*}{\(\textrm{I}_{b_{\textrm{lim}}}\) (p.u.)}
& \multicolumn{2}{c}{Type-I}
& \multicolumn{2}{c}{Type-II} \\
\cmidrule(lr){2-3} \cmidrule(lr){4-5}
& \(N_e\) & $err$
& \(N_e\) & $err$ \\
\midrule
5.00 & 0 & 0 & 11 & -1.79\% \\
\bottomrule
\end{tabular}
}
\label{table:SCC_validation_lastrow_Ne_err}
\vspace{-0.2cm}
\end{table}

\begin{figure}[t!]
\centering
\includegraphics[width=1\columnwidth]{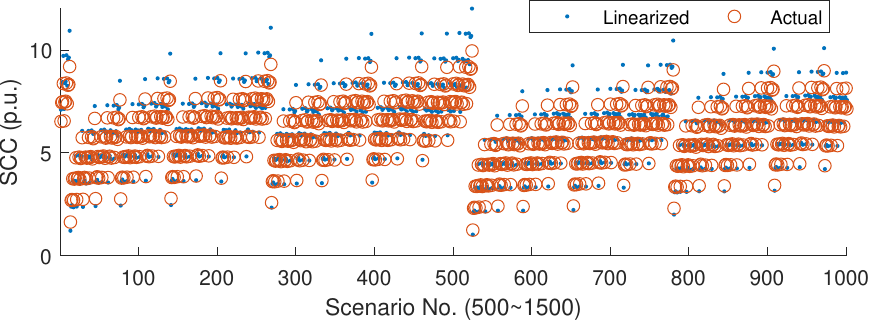}
\vspace{-0.8cm}
\caption{Actual and linearized SCC values for IBR in bus 1.}
\label{SCC_approx}
\vspace{-0.6cm}
\end{figure}

\begin{figure}[t]
\centering
\includegraphics[width=1\columnwidth]{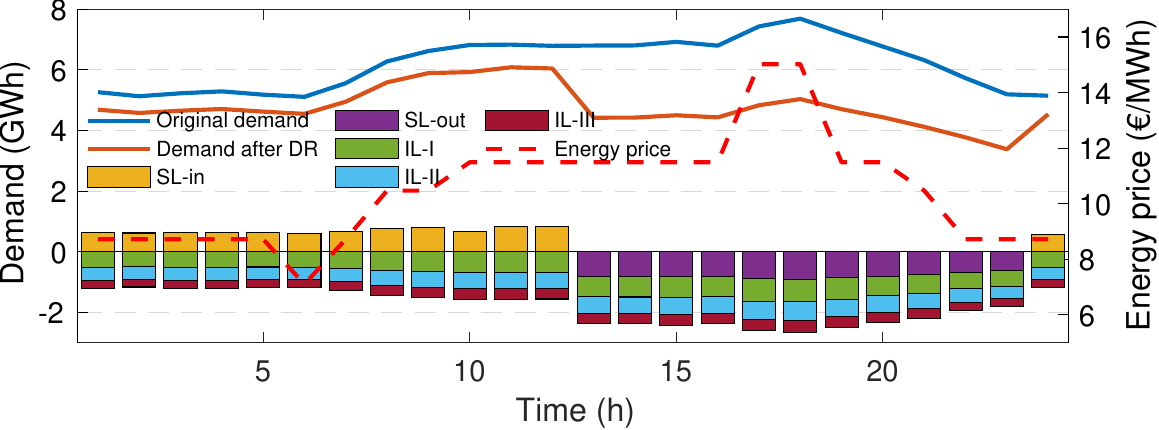}
\vspace{-0.6cm}
\caption{Energy price and system demand in the case of no SCC constraints. SL-in(resp. out) denotes the shifted in(resp. out) flexible load. IL-I, IL-II and IL-III represent three types of IL, respectively.}
\label{Demand_DR}
\vspace{-0.3cm}
\end{figure}

Case studies are conducted on a modified IEEE 30-bus system, as shown in Fig.~\ref{ieee_system}. The IBR, wind turbines, are installed at buses \{1, 23, 26\}, while SGs are located at buses \{2, 3, 4, 5, 27, 30\}, with each bus hosting two SGs. The SCC threshold $\textrm{I}_{b_{\textrm{lim}}}$ for system buses is uniformly set to 5 p.u. to ensure a minimum short-circuit ratio of 3, which is commonly considered indicative of a strong grid. Particularly, the stable operation of IBR buses typically requires a short-circuit ratio of at least 3 \cite{jia2018synchronous}, making this threshold in fact a conservative choice. For other buses, the required SCC level may depend on protection settings and voltage stability margins. In general, determining the minimum SCC is a complex planning task \cite{chu2021short} and should be addressed before the system scheduling; it is therefore beyond the scope of this work. 

The parameters of SGs are listed in Table~\ref{table:SGs_para}, while the parameters of SL and IL (which is classified into three types: I, II, and III) are provided in Table~\ref{table:Parameters Associated with Demand Response}. The notation for SGs is defined as follows: $g_1\textrm{-}b2$ and $g_2\textrm{-}b2$ represent the two individual SGs at bus 2, while 2$g\textrm{-}b2$ stands for the both. The energy price `$\uplambda_t$' is estimated using the pricing method, namely `restricted method' \cite{chu2024pricing}, and treated as a constant in the dispatch. All remaining system parameters are adopted from \cite{wang2025imperfect}. All code and data are publicly available at the repository \cite{Code}. The numerical tests were performed using \texttt{Julia-JuMP} and \texttt{Gurobi-12.0.0}.

\subsection{Validation of Approximate SCC Constraints}
The approximate SCC constraint \eqref{eq:SCC_constraints} is numerically assessed by examining two types of errors \cite{chu2021short}. Type-I: $I_{b_{\textrm{L}}}\geq \textrm{I}_{b_{\textrm{lim}}} \cap I_{{b_{\textrm{SC}}}}<\textrm{I}_{b_{\textrm{lim}}}$, indicates that the approximation is looser than the actual requirement and may lead to insecure operation; Type-II: $I_{b_{\textrm{L}}}<\textrm{I}_{b_{\textrm{lim}}}\cap I_{{b_{\textrm{SC}}}}\geq \textrm{I}_{b_{\textrm{lim}}}$, implies that the approximation is more conservative than the grid-code threshold and thus maintains system security. Since Type-I errors may result in unsafe operating conditions, they are considered undesirable. The averaged value of the errors is defined below:
\begin{equation}\label{eq:error}
 N_e=| \xi |,~   err = \frac{1}{N_e} \sum_{\xi} \frac{I^{(\xi)}_{b_\textrm{L}} - I^{(\xi)}_{b_\textrm{SC}}}{I^{(\xi)}_{b_\textrm{SC}}}
\end{equation}
where $\xi$ indicates the enumerated system operating points. $N_e$ is the total number of errors.

The assessment result is summarized in Table \ref{table:SCC_validation_lastrow_Ne_err}, showing that the approximation eliminates Type-I errors and produces only a small amount of
Type-II errors, indicating a good performance. The approximation result for SCC at IBR in bus 1 is graphically given as Fig.~\ref{SCC_approx} (note that the linearization for other buses is not given due to their similar trend). It can be seen that most actual points have been captured by the approximate ones, showing a desired training performance.

\subsection{System Operation without SCC Constraints}

\begin{table}[t]
\centering
\caption{Costs of SGs and Consumers under Various System Conditions}
\vspace{-0.2cm}
\setlength{\tabcolsep}{2.3pt}
{\fontsize{8pt}{12pt}\selectfont
\begin{tabular}{lccc}
\toprule
Costs (million\texteuro/day) & \makecell{w/o DR,\\ w/o SCC con} & \makecell{with DR,\\ w/o SCC con} & \makecell{with DR,\\ with SCC con}  \\
\midrule
System operation cost               & 1.23  & 0.86 & 0.94 \\
Payment for consumers \eqref{eq:DR_compensation} & 37.29 & 26.01 & 26.01 \\
Total cost \eqref{eq:primal_obj}    & 38.52 & 26.87 & 26.95 \\
\bottomrule
\end{tabular}
}
\label{table:System Operation Cost and Energy Cost}
\vspace{-0.3cm}
\end{table}

\begin{figure}[t]
\centering
\includegraphics[width=1\columnwidth]{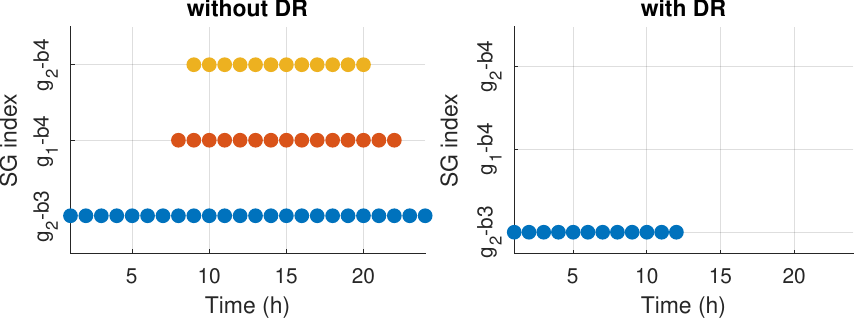}
\vspace{-0.8cm}
\caption{Generators whose commitment state is affected by DR.}
\label{UC_DR}
\vspace{-0.2cm}
\end{figure}

\begin{figure}[t]
\centering
\includegraphics[width=1\columnwidth]{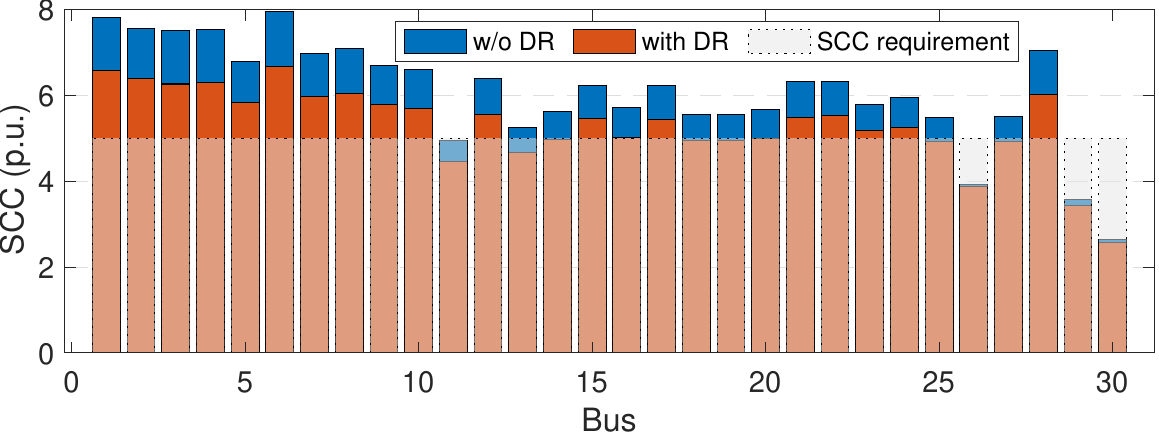}
\vspace{-0.8cm}
\caption{Unconstrained minimum SCC level at each bus.}
\vspace{-0.6cm}
\label{ED_(no)DR}
\end{figure}

\begin{table}[t!]
\centering
\caption{Buses with Inadequate SCC Level}
\vspace{-0.2cm}
{\fontsize{8pt}{12pt}\selectfont
\begin{tabular}{lc}
\toprule
DR implementation & Bus index  \\
\midrule
without DR               & 11, 26, 29, 30    \\
with DR                  & 11, \textbf{13}, \textbf{14}, \textbf{18}, \textbf{19}, \textbf{20}, \textbf{25}, 26, \textbf{27}, 29, 30  \\
\bottomrule
\end{tabular}
}
\label{table:Buses with SCC Inadequacy}
\vspace{-0.4cm}
\end{table}

In this subsection, the SCC constraint is removed from the dispatch \eqref{eq:primal_model}, resulting in a system without SCC security. The energy demand after implementing DR is shown in Fig.~\ref{Demand_DR}. As observed, the users' response leads to a load reduction (from an average of 6.21 GWh to 4.79 GWh), primarily because IL tends to be curtailed to reduce energy charges while receiving compensation. In addition, SL during peak periods with relatively high energy price (13:00–21:00) is shifted to off-peak periods to further minimize energy payments.

The system operation cost and the bill for consumers are given in Table~\ref{table:System Operation Cost and Energy Cost}. Compared with the case without DR, the operation cost of SGs decreases by 30.08\% (from 1.23 m\texteuro~to 0.86 m\texteuro), and the total cost decreases by 30.24\%, mainly due to users’ adjustment of power consumption (with their cost dropping from 37.29 m\texteuro~to 26.01 m\texteuro). This demonstrates the effectiveness of DR in improving the economic efficiency of system operation. However, it comes at the expense of reducing commitments of certain SGs, as depicted in Fig.~\ref{UC_DR}. 

As discussed earlier, inadequate SCC levels may result from few commitments of SGs. Therefore, we calculate the SCC at each bus over the day and assess whether each bus meets the security requirement by comparing its minimum SCC level with the threshold $\textrm{I}_{b_{\textrm{lim}}}$. Based on the obtained UC schedule and \eqref{eq:SCC_constraints}, the minimum SCC at each bus is illustrated in Fig.~\ref{ED_(no)DR}, and Table~\ref{table:Buses with SCC Inadequacy} lists the buses with insufficient SCC. For the case without DR, buses \{11, 26, 29, 30\} have failed to meet the requirements of the protection devices. Specifically, buses \{11, 29\} have no generation units installed and thus rely entirely on SCC supplied by other generating units; in addition, their electrical distance from the lowest-cost generators (i.e., 2$g\textrm{-}b2$, 2$g\textrm{-}b3$, which are typically online) makes them hardly absorb adequate SCC. Buses \{26, 30\} also receive insufficient local SCC support: bus 26 has only one wind turbine with very limited current injection, while bus 30 has two SGs that usually remain offline due to their high generation costs. However, after the implementation of DR, due to the reduced electricity demand, generator $g_2\textrm{-}b3$ is less dispatched and 2$g\textrm{-}b4$ are even shut down over the day (as shown in Fig.~\ref{UC_DR}), resulting in insufficient SCC at additional buses \{13, 14, 18, 19, 20, 25, 27\}, as summarized in Table~\ref{table:Buses with SCC Inadequacy}. 

These results highlight that committed SGs do help ensure the required SCC level at buses; particularly when the load level decreases, it becomes even more crucial to include SCC constraints in the dispatch to prevent a large number of SGs from being offline.

\subsection{SCC Constrained System Operation}

\begin{figure}[t!]
\centering
\includegraphics[width=1\columnwidth]{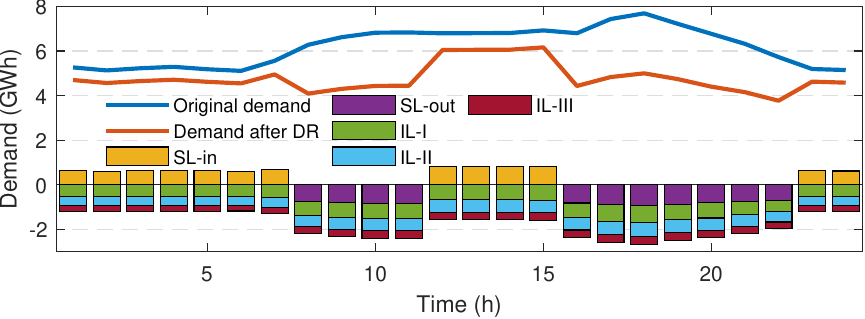}
\vspace{-0.8cm}
\caption{System demand with DR and SCC constraints.}
\label{DR_SCC}
\vspace{-0.2cm}
\end{figure}

\begin{figure}[t!]
\centering
\includegraphics[width=1\columnwidth]{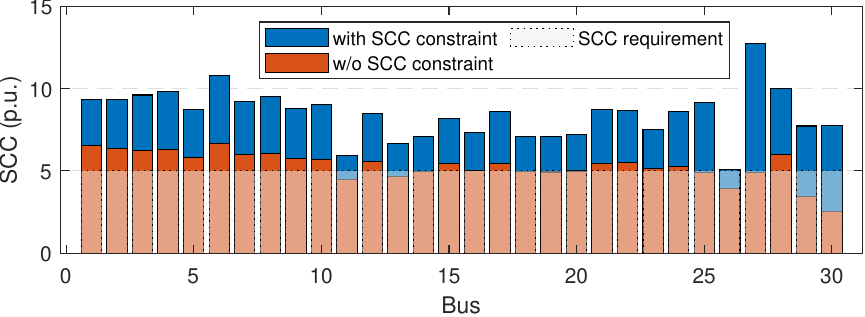}
\vspace{-0.8cm}
\caption{Constrained minimum SCC level at each bus.}
\label{SCC_DR}
\vspace{-0.6cm}
\end{figure}

This subsection applies the complete dispatch model \eqref{eq:primal_model}, enforcing SCC security at all buses. The DR outcomes are shown in Fig.~\ref{DR_SCC}, where the average load level remains nearly unchanged compared with the case without SCC constraints (as seen in Fig.~\ref{Demand_DR}), since the total load curtailment is the same, at 34,063 MWh. To make each bus satisfy the SCC constraint, SGs are dispatched more frequently to provide the needed SCC (as illustrated in Fig.~\ref{SCC_DR}), resulting in a 9.30\% increase in system operation cost (from 0.86 m\texteuro~to 0.94 m\texteuro). The payment for consumers has slightly reduced by 4,457 \texteuro~(0.017\%), as the quantity of responded SL rises by 89 MW. Moreover, part of the SL is shifted to SCC-insufficient periods (13:00–15:00), which also correspond to off-peak energy price hours, thereby enabling more commitments of SGs during these periods for SCC issues without increasing more consumer costs.

Consequently, the system has still achieved a low-cost operation through the coordination of flexible loads and SGs' commitments, with the total cost increasing minorly from 26.87 m\texteuro~to 26.95 m\texteuro~(only 0.3\%), which remains lower than that in the case without DR and SCC constraints (38.52 m\texteuro~in Table~\ref{table:System Operation Cost and Energy Cost}). These results not only emphasize the necessity of SCC constraints but also prove that DR mechanism is conducive to maintaining a low-cost and stable system operation.

\section{Conclusion}\label{Conclusion}
This paper has investigated the impact of DR on SCC levels using a UC model that incorporates both SCC and incentive-based DR constraints. The case results show that the DR mechanism can significantly reduce total costs, although this reduction is achieved primarily through substantial load curtailment at certain periods. Hence, some SGs that were originally online could be shut down at these periods due to the reduced power demand, leading to lower SCC levels at buses and thus rendering protective devices ineffective. When the SCC level at each bus is constrained to meet the safety threshold, the system is able to dispatch flexible loads in a cost-effective manner, ensuring that SGs remain online during SCC-insufficient periods while maintaining the overall cost at an acceptable level. In conclusion, DR can serve as an effective means to facilitate achieving a low-cost system operation when SCC levels are secured.  

In the future, we will investigate more system stability issues caused by market mechanisms, in order to promote the formation of a cost-effective grid with stability secured.


\IEEEtriggeratref{20}
\bibliographystyle{IEEEtran} 
\bibliography{main}

\end{document}